\documentclass[10pt,conference]{IEEEtran}
\usepackage{mathpazo}
\usepackage{times}

\usepackage{amsmath}
\usepackage{amsfonts}
\usepackage{latexsym}
\usepackage{amssymb}

\usepackage{upref}
\usepackage{theorem}
\usepackage{graphicx}
\usepackage{psfrag}
\usepackage{algorithmic}
\usepackage{algorithm}
\usepackage{verbatim}
\usepackage{cite}
\usepackage{color}
\theoremstyle{plain}
\theorembodyfont{\normalfont\slshape}

\newtheorem{thm}{Theorem$\!$}
\newenvironment{theorem}
{\begin{thm}\hspace*{-1ex}{\bf.}}{\end{thm}}

\newtheorem{lem}[thm]{Lemma$\!$}
\newenvironment{lemma}{\begin{lem}\hspace*{-1ex}{\bf.}}{\end{lem}}

\newtheorem{prop}[thm]{Proposition$\!$}

\newtheorem{cor}[thm]{Corollary$\!$}
\newenvironment{corollary}{\begin{cor}\hspace*{-1ex}{\bf.}}{\end{cor}}

\newtheorem{defn}[thm]{Definition$\!$}
\newenvironment{definition}{\begin{defn}\hspace*{-1ex}{\bf.}}{\end{defn}}

\newtheorem{xmpl}[thm]{Example$\!$}
\newenvironment{example}{\begin{xmpl}\hspace*{-1ex}{\bf.}}{\hfill $\Box$ \end{xmpl}}

\newtheorem{cnstr}{Construction$\!$}
\newenvironment{construction}{\begin{cnstr}\hspace*{-1ex}{\bf.}}{\end{cnstr}}

\newcounter{enumrom}
\renewcommand{\theenumrom}{(\roman{enumrom})}

\makeatletter
\renewcommand{\@endtheorem}{\endtrivlist}
\makeatother

\makeatletter
\renewcommand{\thefigure}{{\@arabic\c@figure}}
\renewcommand{\fnum@figure}{{\bf Figure\,\thefigure}}
\makeatother

\newcommand{\cR}{\mathcal{R}}

\newcommand{\mathset}[1]{\left\{#1\right\}}
\newcommand{\abs}[1]{\left|#1\right|}
\newcommand{\ceilenv}[1]{\left\lceil #1 \right\rceil}
\newcommand{\floorenv}[1]{\left\lfloor #1 \right\rfloor}
\newcommand{\parenv}[1]{\left( #1 \right)}

\newcommand{\be}[1]{\begin{equation}\label{#1}}
\newcommand{\ee}{\end{equation}}

\renewcommand{\leq}{\leqslant}

\renewcommand{\geq}{\geqslant}

\renewcommand{\Bbb}{\mathbb}

\newcommand{\Cref}[1]{Co\-ro\-lla\-ry\,\ref{#1}}

\renewcommand{\Bbb}{\mathbb}

\newcommand{\R}{{\Bbb R}}
\newcommand{\Z}{{\Bbb Z}}

\DeclareMathAlphabet{\mathbfsl}{OT1}{cmr}{bx}{it}

\outer\def\proclaim #1. #2\par{\medbreak
 \noindent{\bf#1.\enspace}{\sl#2\par}%
 \ifdim\lastskip<\medskipamount \removelastskip\penalty55\medskip\fi}

\mathchardef\inn="3232
\renewcommand{\in}{{\,\inn\,}}

\newcommand{\bmbf}{\bar{\mathbf{f}}}
\newcommand{\barf}{\bar{f}}
\newcommand{\mbc}{\mathbf{c}}
\newcommand{\mbf}{\mathbf{f}}
\newcommand{\cRb}{\bar{\cR}}

\DeclareMathOperator{\lcm}{lcm}

\newcommand{\bz}{\underline{0}}
\newcommand{\bo}{\underline{1}}

\newcommand{\mbo}{\mathbf{1}}
\newcommand{\mbz}{\mathbf{0}}

\begin{document}

\IEEEoverridecommandlockouts % to enable \thanks

\title{\Huge\bf Generalized Gray Codes for \\ Local Rank Modulation}

\author{
\IEEEauthorblockN{\textbf{Eyal En Gad}}
\IEEEauthorblockA{Elec.~Eng.  \\
Caltech \\
Pasadena, CA 91125, U.S.A. \\
{\it eengad@caltech.edu}\vspace*{-4.0ex}}
\and
\IEEEauthorblockN{\textbf{Michael Langberg}}
\IEEEauthorblockA{Comp.~Sci.~Division  \\
Open University of Israel \\
Raanana 43107, Israel \\
{\it mikel@openu.ac.il}\vspace*{-4.0ex}}
\and
\IEEEauthorblockN{\textbf{Moshe Schwartz}}
\IEEEauthorblockA{Elec.~and Comp.~Eng.  \\
Ben-Gurion University \\
Beer Sheva 84105, Israel \\
{\it schwartz@ee.bgu.ac.il}\vspace*{-4.0ex}}
\and
\IEEEauthorblockN{\textbf{Jehoshua Bruck}}
\IEEEauthorblockA{Elec.~Eng.  \\
Caltech \\
Pasadena, CA 91125, U.S.A. \\
{\it bruck@paradise.caltech.edu}\vspace*{-4.0ex}}
\thanks{
  This work was supported in part by ISF grant 134/10,
  ISF grant 480/08,
  the Open University of Israel's research fund
  (grant no.~46114), the NSF grant ECCS-0802107, and
  an NSF-NRI award.}
}

\maketitle

%%%%%%%%%%%%%%%%%%%%%%%%%%%%%%%%%%%%%%%%%%%%%%%%%%%%%%%%%%%%%%%%%%%%%%%%

\begin{abstract}
We consider the local rank-modulation scheme in which a sliding window
going over a sequence of real-valued variables induces a sequence
of permutations. Local rank-modulation is a generalization of
the rank-modulation scheme, which has been recently suggested as a way of storing information in flash memory.

We study Gray codes for the local rank-modulation scheme
in order to simulate conventional multi-level flash cells while retaining
the benefits of rank modulation. Unlike the limited scope of previous
works, we consider code constructions for the
entire range of parameters including the code length, sliding window size,
and overlap between adjacent windows. We show our constructed codes have
asymptotically-optimal rate. We also provide efficient
encoding, decoding, and next-state algorithms.
\end{abstract}

\section{Introduction}
\label{sec:introduction}

With the recent application to flash memories, the rank-modulation
scheme has gained renewed interest as evident in the recent series of
papers
\cite{JiaMatSchBru09,JiaSchBru10,WanJiaBru09,TamSch10,Sch10,EngLanSchBru10}.
In the conventional modulation scheme used in flash-memory cells, the
absolute charge level of each cell is quantized to one of $q$ levels,
resulting in a single demodulated symbol from an alphabet of size $q$.
In contrast, in the rank modulation scheme a group of $n$ flash cells
comprise a single virtual cell storing a symbol from an alphabet of
size $n!$, where each symbol is assigned a distinct configuration of
relative charge levels in the $n$ cells. Thus, there is no more need
for threshold values to distinguish between various stored symbols,
which mitigates the effects of retention in flash cells (slow charge
leakage).  In addition, if we allow only a simple programming
(charge-injection) mechanism called ``push-to-the-top'', whereby a
single cell is driven above all others in terms of charge level, then
no over-programming can occur, a problem which considerably slows
down programming in conventional multi-level flash cells.

Rank modulation has been studied intermittently since the early works
of Slepian \cite{Sle65} (later extended in \cite{BerJelWol72}), in
which permutations were used to digitize vectors from a time-discrete
memoryless Gaussian source, and Chadwick and Kurz \cite{ChaKur69}, in
which permutations were used in the context of signal detection over
channels with non-Gaussian noise (especially impulse noise).  Other
works on the subject include \cite{ChaRee70, BerJelWol72, Bla74,
  CohDez77, DezFra77, BlaCohDez79}.  More recently, permutations were
used for communicating over powerlines (for example, see
\cite{VinHaeWad00}), and for modulation schemes for flash memory
\cite{JiaMatSchBru09,JiaSchBru10,WanJiaBru09,TamSch10}.

One drawback to the rank-modulation scheme is the fact that we need to
reconstruct the permutation induced by the relative charge levels of
the participating cells.  If $n$ cells are involved, at least
$\Omega(n\log n)$ comparisons are needed, which might be too high for
some applications. It was therefore suggested in
\cite{WanJiaBru09,Sch10,EngLanSchBru10} that only \emph{local}
comparisons be made, creating a sequence of small induced permutations
instead of a single all-encompassing permutation. This obviously
restricts the number of distinct configurations, and thus, reduces
the size of the resulting alphabet as well. In the simplest case,
requiring the least amount of comparisons, the cells are located in a
one-dimensional array and each cell is compared with its two immediate
neighbors requiring a single comparator between every two adjacent
cells \cite{Sch10,EngLanSchBru10}.

Yet another drawback of the rank-modulation scheme is the fact that
distinct $n$ charge levels are required for a group of $n$ physical
flash cells.  Therefore, restricted reading resolution prohibits the
use of large values of $n$. However, when only local views are considered,
distinct values are required only within a small local set of cells,
thus enabling the use of large groups of cells with local rank modulation.

\begin{figure*}[th]
\psfrag{c1}{\scriptsize $c_0=5.00$}
\psfrag{c2}{\scriptsize $c_1=2.50$}
\psfrag{c3}{\scriptsize $c_2=4.25$}
\psfrag{c4}{\scriptsize $c_3=6.50$}
\psfrag{c5}{\scriptsize $c_4=4.00$}
\psfrag{c6}{\scriptsize $c_5=1.00$}
\psfrag{c7}{\scriptsize $c_6=1.50$}
\psfrag{c8}{\scriptsize $c_7=5.50$}
\psfrag{c9}{\scriptsize $c_8=6.00$}
\psfrag{ppp}{$\mbf_\mbc=([3,0,2,4,1],[4,2,0,1,3],[0,3,4,2,1])$}
\hspace{-0.5in}
\includegraphics[scale=0.75]{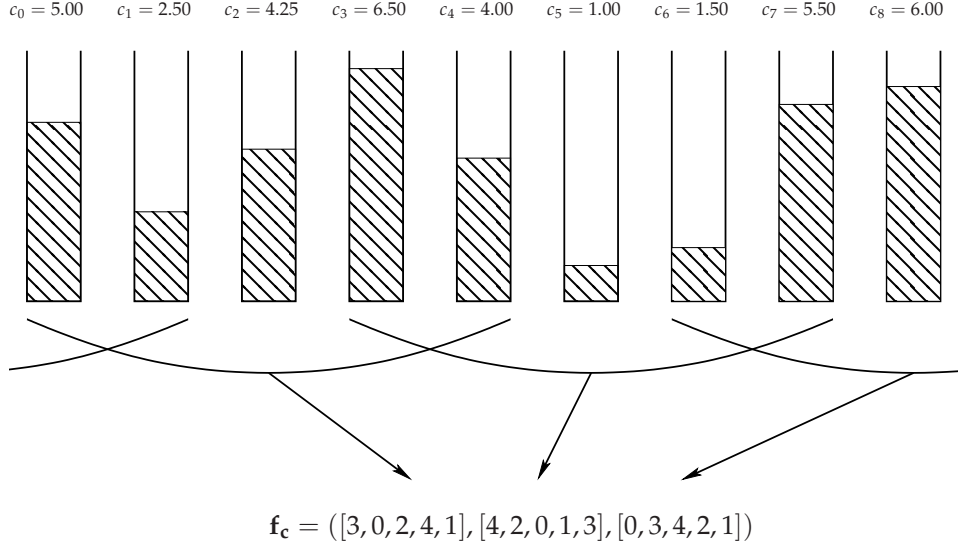}
\caption{Demodulating a $(3,5,9)$-locally rank-modulated signal.}
\label{fig:example}
\end{figure*}

An important application for rank-modulation in the context of flash
memory was described in \cite{JiaMatSchBru09}: A set of $n$ cells,
over which the rank-modulation scheme is applied, is used to simulate
a single conventional multi-level flash cell with $n!$ levels
corresponding to the alphabet $\mathset{0,1,\dots,n!-1}$. The
simulated cell supports an operation which raises its value by $1$
modulo $n!$. This is the only required operation in many rewriting
schemes for flash memories (see
\cite{JiaBohBru07,JiaBru08,YaaSieWol08,BohJiaBru07,JiaLanSchBru09}),
and it is realized in \cite{JiaMatSchBru09} by a Gray code traversing
the $n!$ states where, physically, the transition between two adjacent
states in the Gray code is achieved by using a single
``push-to-the-top'' operation. In the context of local rank modulation,
Gray codes for the local scheme were studied in \cite{Sch10,EngLanSchBru10},
where necessary conditions as well as constructions were provided.

Having considered the two extremes: full rank modulation with a single
permutation of $n$ cells, and extreme local rank modulation with a
sequence of $n$ permutations over $2$ elements, the question of
whether any middle-road solutions exist remains open. We address this
question in this paper by considering the generalized local rank
modulation scheme in which a sequence of several permutations of a
given size provide the local views into ranking of the cells. We
construct Gray codes for this scheme which asymptotically achieve the
maximum possible rate, and consider efficient encoding/decoding algorithms, as
well as efficient next-state computation.

The rest of the paper is organized as follows.
In Section~\ref{sec:notation} we give preliminary definitions and notation.
In Section~\ref{sec:construction} we present our construction for optimal local rank modulation for general degrees of locality.
We conclude with a discussion in Section~\ref{sec:conclusion}.

%%%%%%%%%%%%%%%%%%%%%%%%%%%%%%%%%%%%%%%%%%%%%%%%%%%%%%%%%%%%%%%%%%%%%%%%
%%%%%%%%%%%%%%%%%%%%%%%%%%%%%%%%%%%%%%%%%%%%%%%%%%%%%%%%%%%%%%%%%%%%%%%%
%%%%%%%%%%%%%%%%%%%%%%%%%%%%%%%%%%%%%%%%%%%%%%%%%%%%%%%%%%%%%%%%%%%%%%%%

\section{Definitions and Notation}
\label{sec:notation}

We shall now proceed to introduce the notation and definitions pertaining
to local rank modulation and Gray codes.
We will generally follow the notation introduced in \cite{Sch10,EngLanSchBru10}.

\subsection{Local Rank Modulation}

Let us consider a sequence of $t$ real-valued variables,
$\mbc=(c_0,c_1,\dots,c_{t-1})\in \R^t$, where we further assume
$c_i\neq c_j$ for all $i\neq j$.  The $t$ variables induce a
permutation $f_\mbc\in S_t$, where $S_t$ denotes the set of all
permutations over $[t]=\mathset{0,1,2,\dots,t-1}$.  The permutation
$f_\mbc$ is defined as
\[f_\mbc(i)=\abs{\mathset{j ~|~ c_j < c_i}}.\]
Loosely speaking, $f_\mbc(i)$ is the rank of the $i$th cell in ascending
order. This ranking is equivalent to the permutation
described in \cite{Sch10,EngLanSchBru10}, though different.

Given a sequence of $n$ variables, $\mbc=(c_0,c_1,\dots,c_{n-1})$, we define a
window of size $t$ at position $p$ to be
\[\mbc_{p,t}=(c_p,c_{p+1},\dots,c_{p+t-1})\]
where the indices are taken modulo $n$, and also $0\leq p\leq n-1$,
and $1\leq t\leq n$.
We now define the \emph{(s,t,n)-local rank-modulation (LRM) scheme}, which we do
by defining the \emph{demodulation} process.
Let $s\leq t\leq n$ be positive integers, with $s|n$.
Given a sequence of
$n$ distinct real-valued variables, $\mbc=(c_0,c_1,\dots,c_{n-1})$,
the demodulation maps $\mbc$ to the sequence of $n/s$ permutations from $S_t$
as follows:
\begin{equation}
\label{eq:seqper}
\mbf_\mbc=(f_{\mbc_{0,t}},f_{\mbc_{s,t}},f_{\mbc_{2s,t}},\dots,f_{\mbc_{n-s,t}}).
\end{equation}
Loosely speaking, we scan the $n$ variables using
windows of size $t$ positioned at multiples of $s$ and write down the
permutations from $S_t$ induced by the \emph{local} views of the sequence.

In the context of flash-memory storage devices, we shall consider the
$n$ variables, $\mbc=(c_0,c_1,\dots,c_{n-1})$, to be the charge-level
readings from $n$ flash cells. The demodulated sequence, $\mbf_\mbc$, will
stand for the original information which was stored in the $n$ cells.
This approach will serve as the main motivation for this paper, as it was
also for \cite{JiaMatSchBru09,JiaSchBru10,TamSch10,WanJiaBru09,Sch10,EngLanSchBru10}.
See Figure \ref{fig:example} for an example.

We say a sequence $\mbf$ of $n/s$ permutations over $S_t$ is
\emph{$(s,t,n)$-LRM realizable} if there exists $\mbc\in\R^n$ such
that $\mbf=\mbf_{\mbc}$, i.e., it is the demodulated sequence of
$\mbc$ under the $(s,t,n)$-LRM scheme. Except for the degenerate case
of $s=t$, not every sequence is realizable. We denote the set of all
$(s,t,n)$-LRM realizable permutation sequences as $\cR(s,t,n)$.
In a later part of this section, we show that the number of states representable by an $(s,t,n)$-LRM scheme, i.e., the size of $\cR(s,t,n)$, is roughly
$(t \cdot(t-1) \cdot ... \cdot (t-s+1))^{n/s}$  (this fact is also stated in \cite{WanJiaBru09}).

While any $\mbf\in\cR(s,t,n)$ may be represented as a sequence of
$n/s$ permutations over $S_t$, a more succinct representation is
possible based on the (mixed-radix) factoradic notation system (see
\cite{Lai88} for the earliest-known definition, and
\cite{JiaMatSchBru09} for a related use): We can represent any
permutation $f=[f(0),\dots,f(t-1)]\in S_t$ with a sequence of digits
$d_{t-1},d_{t-2},\dots,d_1,d_0$, where $d_i\in\Z_i$, and $d_i$ counts the
number of elements to the right of $f(i)$ which are of lower value. We call
$d_{t-1}$ the \emph{most significant digit} and $d_0$ the \emph{least significant
digit}. If
$f=f_\mbc$ for some $\mbc\in\R^t$, then the factoradic representation is easily seen to be
equivalent to counting the number of cells to the right of the $i$th cell
which are with lower charge levels.

Continuing with our succinct representation, we now contend that due
to the overlap between local views, we can then represent each of the
local permutations $f_{\mbc_{i\cdot s,t}}$ using only the $s$
most-significant digits in their factoradic notation.  We denote this
(partial) representation as $\barf_{\mbc_{i\cdot s,t}}$. Accordingly, we define,
\[
\bmbf_\mbc= (\barf_{\mbc_{0,t}},\barf_{\mbc_{s,t}},\barf_{\mbc_{2s,t}},\dots,\barf_{\mbc_{n-s,t}}),
\]
and the set of all such presentations as $\cRb(s,t,n)$.
Thus, for example, the configuration
of Figure \ref{fig:example} would be represented by
$\parenv{(3,0,1),(4,2,0),(0,2,2)}$.

\begin{lemma}
\label{lemma:biject}
For all $1\leq s \leq t \leq n$,
\[\abs{\cRb(s,t,n)}\leq \abs{\cR(s,t,n)}\leq (t-s)!\cdot\parenv{\frac{t!}{(t-s)!}}^{\frac{n}{s}}.\]
\end{lemma}

\begin{IEEEproof}
That $\abs{\cRb(s,t,n)}\leq \abs{\cR(s,t,n)}$ is trivial, since any $\mbf\in\cR(s,t,n)$ results in one
$\bmbf\in\cRb(s,t,n)$. For the other inequality, assume we fix the permutation induced by
the first  $t-s$ cells, where there are $(t-s)!$ ways of doing so. It follows that there are
$t!/(t-s)!$ ways of choosing $f_{\mbc_{n-s,t}}$, and then the same amount of ways of choosing
$f_{\mbc_{n-2s,t}}$, and continuing all the way up to $f_{\mbc_{0,t}}$ we get the desired result.
\end{IEEEproof}

When $s=t=n$, the $(n,n,n)$-LRM scheme degenerates
into a single
permutation from $S_n$.
This was the case in most of the previous works using permutations for
modulation purposes.
A slightly more general case, $s=t < n$ was discussed by Ferreira
\emph{et al.} \cite{FerVinSwaBee05} in the context of permutation trellis
codes, where a binary codeword was translated tuple-wise into a sequence
of permutation with no overlap between the tuples. An even more
general case was defined by Wang \emph{et al.} \cite{WanJiaBru09} (though
in a slightly different manner where indices are not taken modulo $n$, i.e.,
with no wrap-around). In \cite{WanJiaBru09}, the sequence of permutations
was studied under a charge-difference constraint called
\emph{bounded rank-modulation}, and mostly with parameters
$s=t-1$, i.e., an overlap of one position between adjacent windows.
Finally, using the same terminology as this paper, the case of $(1,2,n)$-LRM
was considered in \cite{Sch10,EngLanSchBru10}.

\subsection{Gray Codes}

Generally speaking, a \emph{Gray code}, $G$, is a sequence of distinct states
(codewords),
$G=g_0,g_1,\dots,g_{N-1}$, from an ambient state space, $g_i\in S$, such that
adjacent states in the sequence differ by a ``small'' change. What constitutes
a ``small'' change usually depends on the code's application.

Since we are interested in building Gray codes for flash memory
devices with the $(s,t,n)$-LRM scheme, our ambient space is
$\cR(s,t,n)$, which is the set of all realizable sequences under
$(s,t,n)$-LRM.

The transition between adjacent states in the Gray code is directly
motivated by the flash memory application, and was first described
and used in \cite{JiaMatSchBru09}, and later also used in
\cite{Sch10,EngLanSchBru10}. This transition is the ``push-to-the-top''
operation, which takes a single flash cell and raises its charge level
above all others.

In our case, however, since we are considering a \emph{local}
rank-modulation scheme, the ``push-to-the-top'' operation merely
raises the charge level of the selected cell above those cells which
are comparable with it.  Thus, we define the set of allowed
transitions as $T=\mathset{\tau_0,\tau_1,\dots,\tau_{n-1}}$, which is
a set of functions, $\tau_j:\cR(s,t,n)\rightarrow \cR(s,t,n)$, where
$\tau_j$ represents a ``push-to-the-top'' operation performed on the
$j$-th cell. More precisely, let $\mbf$ be an $(s,t,n)$-LRM realizable
sequence of permutations, i.e., there exists $\mbc\in\R^n$ such that
$\mbf=\mbf_\mbc$. Now define the transition $\tau_j$ acting on $\mbf$
as $\mbf'=\mbf'_{\mbc'}$ realizable by the variables $\mbc'=(c'_0,\dots,c'_{n-1})\in\R^n$ such that $c'_j$ is pushed to a value higher than all of it's comparable cells. We denote $r(j)$ as the rightmost index (cyclically) among the cells that share a window with $c'_j$, and $l(j)$ as the leftmost index (cyclically) among them. We can find $r(j)$ and $l(j)$ by the following expressions:
\begin{align*}
l(j) & = s\left\lceil{\frac{j-t+1}{s}}\right\rceil \mod n, \\
r(j) & = \parenv{s\floorenv{\frac{j}{s}} + (t-1)} \mod n.
\end{align*}
Now $\mbc'$ is given by the following expression:
\[
c'_i = \begin{cases}
c_i & i\neq j, \\
\max\mathset{c_{l(i)},\dots,c_{r(i)}}+1 & i=j.
\end{cases}
\]

\begin{definition}
A \emph{Gray code $G$ for $(s,t,n)$-LRM (denoted $(s,t,n)$-LRMGC)}
is a sequence of distinct codewords, $G=g_0,g_1,\dots,g_{N-1}$, where
$g_i\in \cR(s,t,n)$.  For all $0\leq i \leq N-2$, we further require that
$g_{i+1}=\tau_j(g_i)$ for some $j$.  If $g_0=\tau_j(g_{N-1})$ for some
$j$, then we say the code is \emph{cyclic}. We call $N$ the
\emph{size} of the code, and say $G$ is \emph{optimal} if
$N=\abs{\cR(s,t,n)}$.
\end{definition}

\begin{definition}
%Let $G$ be a $(s,t,n)$-LRMGC of size $N$.
%We define the \emph{rate} of the code as $\rate(G)=\frac{\log_2 N}{n}$.
We say a family of codes, $\mathset{G_i}_{i=1}^{\infty}$, where $G_i$ is an
$(s,t,n_i)$-LRMGC of size $N_i$, $n_{i+1}>n_i$, is \emph{asymptotically optimal} if
\[\displaystyle\lim_{i\to\infty}\frac{\log_2 N_i}{\log_2\abs{\cR(s,t,n_i)}}=1.\]
\end{definition}

\section{Gray Codes for $(s,t,n)$-LRM}
\label{sec:construction}

In this section we present \emph{efficiently} encodable and decodable \emph{asymptotically-optimal} Gray codes for $(s,t,n)$-LRM.
A rough description of our construction follows.
First we partition the $n$ cells into $m$ blocks each of size $m/n$.
To simplify our presentation we set $m=\sqrt{n}$, implying that we have $m$ blocks, each of size $m$.
Denote the cells in block $i$ by $\mbc_i$.
For each block $\mbc_i$ we will use the factoradic representation $\bmbf_{\mbc_i}$ to represent permutations in $\cRb(s,t,m)$.
Namely, each and every block can be thought of an element of an alphabet $\Sigma=\{v_0,\dots,v_{V-1}\}$ of size $V$.

Now, consider any de-Bruijn sequence $S$ of order $m-1$ over $\Sigma$ (of period $V^{m-1}$).
Namely, $S$ will consist of a sequence of $V^{m-1}$ elements $v_{s_0},v_{s_1},\dots,v_{s_{V^{m-1}-1}}$ over $\Sigma$ such that the subsequences $v_{s_i},\dots,v_{s_{i+m-2}}$ of $S$ \emph{cover} all $(m-1)$-tuples of $\Sigma$ exactly once, sub-indices of $s$ taken modulo $V^{m-1}$.
Here, $s_i \in [V]$.
Such sequences $S$ exist, e.g., \cite{Gol67}.

We are now ready to construct our Gray code $G$.
The construction will have two phases.
First we construct so-called \emph{anchor} elements in $G$, denoted as $\bar{G} = \{g_0,\dots,g_{L-1}\}$.
The elements of $\bar{G}$ will consist of a cyclic Gray code over $\Sigma^m$.
That is, the difference between each $g_i$ and $g_{i+1}$ in $\bar{G}$ will be in only one out of the $m$ characters (from $\Sigma$) in $g_i$.
Specifically, the code $\bar{G}$ will be derived from the de-Bruijn sequence $S$ as follows: we set $g_0$ to be the first $m$ elements of $S$, and in the transition from $g_i$ to $g_{i+1}$ we change
$v_{s_i}$ to $v_{s_{i+m}}$.
The code $\bar{G}$ is detailed below:
\[
\begin{array}{rcccccc}
g_0= & v_{s_{m-1}} & v_{s_{m-2}} & \dots & v_{s_1} & \underline{v_{s_0}}  \\
g_1= & v_{s_{m-1}} & v_{s_{m-2}} & \dots & \underline{v_{s_1}}   & v_{s_{m}} \\
g_2= & v_{s_{m-1}} & v_{s_{m-2}} & \dots   & v_{s_{m+1}}  & v_{s_{m}} \\
     &      &           & \vdots  &           &             \\
%g_{m-1}= & v_{s_{m-1}} & v_{s_{2m-2}} & \dots  & v_{s_{m+1}}  & v_{s_{m}} \\
%g_{m}= & v_{s_{2m-1}} & v_{s_{2m-2}} & \dots  & v_{s_{m+1}}  & v_{s_{m}} \\
%g_{m+1}=  & v_{s_{2m-1}} & v_{s_{2m-2}} & \dots  & v_{s_{m+1}}  & v_{s_{2m}}   \\
%     &      &           & \vdots &           &            \\
%g_{L-m-1}= & v_{s_{L-m-1}} & v_{s_{L-2}} & \dots & v_{s_{L-m-1}} & v_{s_{L-m}}  \\
%g_{L-m}= & v_{s_{L-1}} & v_{s_{L-2}} & \dots & v_{s_{L-m-1}} & v_{s_{L-m}}  \\
%g_{L-m+1}= & v_{s_{L-1}} & v_{s_{L-2}} & \dots & v_{s_{L-m-1}} & v_{s_0}  \\
%     &      &           & \vdots  &           &                     \\
g_{L-2}= & v_{s_{L-1}} & \underline{v_{s_{L-2}}} & \dots & v_{s_1} & v_{s_0}  \\
g_{L-1}=  & \underline{v_{s_{L-1}}} & v_{s_{m-2}} & \dots & v_{s_1} & v_{s_0}  \\
\end{array}
\]
where $L=\lcm(m,V^{m-1})$, the sub-indices of $s$ are taken modulo $V^{m-1}$, and the underline is an
imaginary marking distinguishing the block which is about to change.

With the imaginary marking of the underline, the code $\bar{G}$ is clearly a Gray code over $\Sigma^m$ due to the properties of the de-Bruijn sequence $S$. However $\bar{G}$ does not suffice for our construction as the transitions between the anchors  $g_i$ and $g_{i+1}$ involve changing the entries of an entire block, which may involve many push-to-the-top operations. We thus refine $\bar{G}$ by adding additional elements between each pair of adjacent anchors from $\bar{G}$ that allow us to move from the block configuration in $g_i$ to that in $g_{i+1}$ by a series of push-to-the-top operations.
Our construction is summarized below formally.

\begin{construction}
\label{con:debruijn}
We consider the $(s,t,n)$-LRM, $n$ be a square, $m=\sqrt{n}\geq t+2$, and require that $s|m$. Let $\mathset{v_0,v_1,\dots,v_{V-1}}$ be a set of $V$ distinct mixed-radix vectors of length $m$ taken
from $([t]\times[t-1]\times\dots\times[t-s])^{m/s}$.
The values of the last $s(\ceilenv{(t+2)/s}-1)$ digits of each $v_i$ do not play a role in the representation of the stored data and are called \emph{non-information digits}, so
by abuse of notation, a mixed-radix vector $(r_0,r_1,\dots,r_{m-1})$ actually represents the value
$(r_0,r_1,\dots,r_{m-1-s(\ceilenv{(t+2)/s}-1)})$ regardless of the value of the last $s(\ceilenv{(t+2)/s}-1)$ elements.
Therefore, we get
\[V=\parenv{\frac{t!}{(t-s)!}}^{\frac{m}{s}-\ceilenv{\frac{t+2}{s}}+1}.\]
We also denote $L=\lcm(m,V^{m-1})$.

Consider a de-Bruijn sequence $S$ of order $m-1$ over the alphabet $\mathset{0,1,\dots,V-1}$.
%The sequence is of period $V^{m-1}$ and we denote it by $s_0,s_1,\dots,s_{V^{m-1}}$. We remind the reader that windows of size $m-1$ in the sequence, i.e., $s_{i},s_{i+1},\dots,s_{i+m-2}$, with indices taken modulo $V^{m-1}$, are all distinct. Such sequences can always be constructed (for example, see \cite{Gol67}).
The Gray code $\bar{G}$ of anchor vectors is a sequence $g_0, g_1, \dots, g_{L-1}$ of $L$ mixed-radix vectors of length $m^2=n$. Each vector is formed by a concatenation of $m$ blocks of length $m$.
%We call $g_0,g_1,\dots,g_{L-1}$ the \emph{anchor vectors}.
Between the anchors $g_{i}$ and $g_{i+1}$, the block $v_{s_i}$ is transformed into the block $v_{s_{i+m}}$.

Within each of the $m$ blocks comprising any single anchor, the
$(m-2)$nd digit (the second-from-right digit -- a non-information digit) is set to
$1$ in all blocks except for the underlined block. For brevity, we call
this digit the \emph{underline digit}.

Between any two anchors, $g_i$ and $g_{i+1}$, a sequence of vectors called \emph{auxiliary vectors} and denoted $g_i^0, g_i^1,\dots, g_i^{\ell_i}$, is formed by a sequence of push-to-the-top operations on the cells of the changing block. The auxiliary vectors are determined by Algorithm \ref{newblock} described shortly.
\end{construction}

In what follows we present Algorithm~\ref{newblock} that specifies the sequence $g_i^0, g_i^1,\dots, g_i^{\ell_i}$ that allow us to move from anchor state $g_i$ to state $g_{i+1}$.
As $g_i$ and $g_{i+1}$ differ only in a single block (and this block is changed from $v_{s_i}$ to $v_{s_{i+m}}$), the same will hold for the sequence $g_i^0, g_i^1,\dots, g_i^{\ell_i}$, i.e., $g_i^j$ and $g_i^{j'}$ will only differ in the block in which $g_i$ and $g_{i+1}$ differ.
Thus, it suffices to define in Algorithm~\ref{newblock} how to change a block of length $m$ with cell values that represent $v_{s_i}$ into a block that represents $v_{s_{i+m}}$ using push-to-the-top operations.
However, we call the attention of the reader to the fact that while the change in represented value affects
only one block, for administrative reasons we also push cells of the block to the left
(cyclically).

We now present Algorithm~\ref{newblock} and describe some of its properties. We then prove that indeed the resulting code $G$ is an asymptotically-optimal cyclic $(s,t,n)$-LRMGC. We assume that the following algorithm is
applied to positions $\mathset{0,1,\dots,m-1}$. We further assume
$(r_0,r_1,\dots,r_m)\in ([t]\times[t-1]\times\dots\times[t-s])^{m/s}$ represents the value
$v_\ell$, then we say the $j$th digit of $v_{\ell}$ is
\[v_\ell(j)=\begin{cases}
r_j & 0\leq j < m-s(\ceilenv{(t+2)/s}-1) \\
0 & \text{otherwise.}
\end{cases}\]
Finally, we restrict $l(\cdot)$ and $r(\cdot)$ by defining
\begin{align*}
l'(j) &= \begin{cases}
l(j) & 0\leq l(j)\leq m-3 \\
0 & \text{otherwise}
\end{cases}\\
r'(j) &= \begin{cases}
r(j) & 0\leq r(j)\leq m-3 \\
m-3 & \text{otherwise}
\end{cases}
\end{align*}

%\paragraph{Algorithm~\ref{newblock}}

%among them, and then start changing that block.
%next step is pushing the last cell of the

%The order of the push-to-the-top operations between anchors $g_i$ and $g_{i+1}$ is determined by Algorithm \ref{newblock}, which also takes care of maintaining the value of the first and last digits discussed above. The idea of the algorithm is pushing each cell to the top after a number of cells equal to its desired rank were already being pushed before. Since a cell's rank correspond to the following cells that share a window with it, we compare it's rank among those cells. We consider the desired rand of the last $t$ cells of the block to be $0$. In order to keep track of which cells were pushed, we save an array of bits for each cell in the block, indicating whether the cell been pushed before. We denote that array as $a$, and initial it to zeros.

\begin{algorithm}[ht]
\caption{Transform block $v_{s_i}$ to block $v_{s_{i+m}}$}
\label{newblock}
\begin{algorithmic}
\STATE Push the rightmost cell of the block to the left (cyclically)
\STATE $a_j \Leftarrow 0$ for all $j=0,1,\dots,m-3$
\STATE $j \Leftarrow 0$
\REPEAT
\IF{$v_{s_{i+m}}(j) =	\displaystyle\sum_{i=j + 1}^{r'(j)}a_i$ and $a_j = 0$}
\STATE Push the $j$th cell of current block.
\STATE $a_j \Leftarrow 1$
\STATE $j \Leftarrow l'(j)$
\ELSE
\STATE $j \Leftarrow j + 1$
\ENDIF
\UNTIL{$j = m-2$}
\STATE Push the next-to-last cell of current block.
\end{algorithmic}
\end{algorithm}

Our algorithm changes a block of length $m$ with cell values that represent $v_{s_i}$ into  one that represents $v_{s_{i+m}}$ using push-to-the-top operations. It is strongly based on the factoradic representation of $v_{s_{i+m}}$.
Let $v_{s_{i+m}}(j)$ be the $j$th entry in this representation.
Namely, if $\mbc=(c_1,\dots,c_m)$ is a cell configuration that corresponds to $v_{s_{i+m}}$, then for each index $j \in [m]$ the number of entries in the window corresponding to $j$ that are to the right of $j$ and are of value lower than $c_j$ equal $v_{s_{i+m}}(j)$.
Roughly speaking, to obtain such a configuration $\mbc$, our algorithm, for $j\in [m]$, pushes each cell $c_j$  in $\mbc$ to the top exactly once and only after exactly $j$ cells to the right of $c_j$ (and participating in the window corresponding to $j$) have been pushed to the top.
As each time a cell is changed it is pushed to the top, this will ensure that the resulting cell configuration $\mbc$ will have a factoradic representation corresponding to $v_{s_{i+m}}$.

A few remarks are in place.
In order to keep track of which cells were pushed during our algorithm, we save an array of bits $a_j$ for each cell in the block (initialized to 0), indicating whether the cell $c_j$ has been pushed before.
We note that in order to be able to decode a state, we need to have some way to know which block is being currently changed, i.e., the imaginary underline in the anchor.
We use the last two cells of each block for that purpose.

\begin{example}
Take the case of $(1,2,16)$-LRM with $m=4$, $V=2$, and a de-Bruijn sequence of order $3$ and alphabet of size $2$ is $S=00010111$. The list of anchors is
\[\begin{array}{rcccc}
g_0 = & \mbo010 & \mbz010 & \mbz010 & \underline{\mbz000} \\
g_1 = & \mbo010 & \mbz010 & \underline{\mbz000} & \mbz010 \\
g_2 = & \mbo010 & \underline{\mbz000} & \mbo010 & \mbz010 \\
g_3 = & \underline{\mbo000} & \mbo010 & \mbo010 & \mbz010 \\
g_4 = & \mbo010 & \mbo010 & \mbo010 & \underline{\mbz000} \\
g_5 = & \mbo010 & \mbo010 & \underline{\mbo000} & \mbz010 \\
g_6 = & \mbo010 & \underline{\mbo000} & \mbz010 & \mbz010 \\
g_7 = & \underline{\mbo000} & \mbz010 & \mbz010 & \mbz010
\end{array}\]
The bold bit (the leftmost bit in each group of four) denotes the information bit, while the rest are
non-information bits. We note that the underlined vectors are easily recognizable by next-to-right bit being
$0$.

Notice that in this example the information bit is dominated in size by the remaining bits of each block. This is an artifact of our example in which we take $n$ be be small. For large values of $n$ the overhead in each block is negligible with respect to the information bits.

As an example, the transition between $g_1$ and $g_2$ is
(the changed positions are underlined)
\[\begin{array}{rcccc}
g_1      = & 1010 & 0010       & 0000       & 0010 \\
g_1^{ 0} = & 1010 & 00\bz\bo   & 0000       & 0010 \\
g_1^{ 1} = & 1010 & 0001       & 0\bo00     & 0010 \\
g_1^{ 2} = & 1010 & 000\bz     & \bo100     & 0010 \\
g_2      = & 1010 & 0000       & 1\bz\bo0   & 0010
\end{array}\]
\end{example}

%As we will prove shortly, if only a single block has a last digit with a value lower than it's maximum, then the block to its right is the block being changed.
%If two (cyclically) consecutive blocks have that property, the rightmost block is in the process of being changed.

%The idea of the algorithm is pushing each cell to the top after a number of cells equal to its desired rank were already being pushed before. Since a cell's rank correspond to the following cells that share a window with it, we compare it's rank among those cells. We consider the desired rand of the last $t$ cells of the block to be $0$. In order to keep track of which cells were pushed, we save an array of bits for each cell in the block, indicating whether the cell been pushed before. We denote that array as $a$, and initial it to zeros.

We now address the analysis of Algorithm~\ref{newblock}.

\begin{lemma}
Assuming the underline is known, all anchors used in Construction
\ref{con:debruijn} are distinct.
\end{lemma}

\begin{IEEEproof}
Proof follows directly from the properties of the de-Bruijn sequence $S$ and the fact that we are taking $L$ to be the $\lcm(m,V^{m-1})$.
\end{IEEEproof}

\begin{lemma}
Algorithm \ref{newblock} maintains the correctness of the underline
digit in anchors. In addition, between any two adjacent anchors, Algorithm
\ref{newblock} guarantees the underline digits of the changing block and
the block to its left (cyclically), are both not maximal.
\end{lemma}

\begin{IEEEproof}
Proof is by induction.
The base case follows from our construction of the first anchor element $g_0$.
Assume $g_i$ satisfies the inductive claim.
When applying Algorithm \ref{newblock} to move from anchor $g_i$ to $g_{i+1}$, we start by pushing the rightmost cell of the block to the left of that being changed. This implies that the value of the underline cell in both the block being changed and that to its left are now not maximal.
This state of affairs remains until the end of Algorithm 1, in which we push the next-to-the last cell in the changed block.
At that point in time, the underline cell in the changed block obtains it's maximal value, while the block to its left (that to be changed in the next application of Algorithm \ref{newblock}) is of non-maximal value.
All the other underline cells remain unchanged throughout the execution of Algorithm \ref{newblock}.
\end{IEEEproof}

\begin{lemma}
\label{lem:all_pushed}
Algorithm \ref{newblock} terminates, and when it does, all of the cells are pushed exactly once.
\end{lemma}
\begin{IEEEproof}
That Algorithm \ref{newblock} terminates is easy to see.
For convenience, we denote $z=s(\ceilenv{(t+2)/s}-1)$.
For each $k\in\{m-z,\dots,m-3\}$, $v_{s_{i+m}}(k)=0$, and therefore each of those cells is pushed the first time that $j=k$. Now we assume by induction that for each $k\leq m-z$, all of the cells with indices $j$, $k\leq j\leq m-3$, are pushed before the algorithm terminates.

The base case, $k=m-z$, was already proved above. For the induction step, by the induction assumption, we know that all the cells in $\{k,\dots,r'(k)\}$ are pushed. At the point where exactly $v_{s_{i+m}}(k-1)$ of them are pushed, cell $k-1$ is pushed in the next visit. Since also the algorithm never pushes a cell more than once, the claim is proved.
\end{IEEEproof}

\begin{theorem}
\label{th:alg_correct}
Algorithm \ref{newblock} changes a block representing $v_{s_i}$ into a block representing $v_{s_{i+m}}$.
\end{theorem}

\begin{IEEEproof}
Before cell $j$ is being pushed, exactly $v_{s_{i+m}}(j)$ cells from $\mathset{j+1,\dots,r'(j)}$ have been
pushed already. The rest will be pushed after and above it, and therefore its rank is exactly $v_{s_{i+m}}(j)$, as desired.
\end{IEEEproof}

One drawback of Algorithm \ref{newblock} is that it may visit a codeword multiple times. For example, assume a $(1,2,25)$-LRM scheme, with $v_{s_{i}}=11XXX$ and $v_{s_{i+5}}=10XXX$, where $X$ is the ``don't care'' symbol. The algorithm would, after an initial push of a cell on the adjacent block to the left, first push cell $1$, changing the block state to $01XXX$. Afterwards, the algorithm would push cell $0$, changing the state back to $v_{s_{i}}$.

To solve that problem, we suggest to simulate the entire remaining execution of the algorithm every time we push a cell. If the resulting configuration after the planned push appears another time in the future, we change the algorithm's inner state to that of the latest such repeat appearance. That way we make sure that each codeword appears only once in the Gray code. We call the revised algorithm the \emph{repetition-avoiding algorithm}.

\begin{lemma}
\label{lem:time_comp}
The time complexity of the repetition-avoiding algorithm is $O(tn)$.
\end{lemma}
\begin{IEEEproof}
Each cell is visited by the algorithm at most $t$ times, once during the first visit of the algorithm, and once following each of the $t-1$ cells immediately to its right being pushed. Since each cell is pushed exactly once,
a full execution of the algorithm takes $O(tm)$ steps. For the repetition-avoiding algorithm, simulating a full execution after each push results in
total time complexity $O(tm +tm^2)=O(t n)$.
\end{IEEEproof}

Combining all of our observations up to now, we are able to summarize with the following theorem for $G$ from Construction~\ref{con:debruijn}.

\begin{theorem}
\label{the:end}
$G$ is a cyclic gray code of size at least $L$.
\end{theorem}

\begin{corollary}
For all constants $1\leq s < t$,
there exists an asymptotically-optimal family of codes, $\mathset{G_i}_{i=t+2}^{\infty}$, where $G_i$ is an
$(s,t,n_i)$-LRMGC of size $N_i$, $n_{i+1}>n_i$, with
\[\displaystyle\lim_{i\to\infty}\frac{\log_2 N_i}{\log_2 \abs{\cR(s,t,n_i)}}=1.\]
\end{corollary}
\begin{IEEEproof}
We set $n_i=s^2 i^2$ for all $i\geq t+2$. Then $N_i\geq L_i\geq V_i^{si-1}$. It follows that
\begin{align*}
& \lim_{i\rightarrow\infty}\frac{\log_2 N_i}{\log_2\abs{\cR(s,t,n_i)}} \geq \\
& \quad \geq \lim_{i\rightarrow\infty}\frac{(si-1)\log_2 V_i}{\log_2\parenv{(t-s)!\cdot\parenv{\frac{t!}{(t-s)!}}^{si^2}}} \\
& \quad = \lim_{i\rightarrow\infty}\frac{(si-1)\log_2
\parenv{\frac{t!}{(t-s)!}}^{i-\ceilenv{\frac{t+2}{s}}+1}}
{\log_2\parenv{(t-s)!\cdot\parenv{\frac{t!}{(t-s)!}}^{si^2}}}\\
&\quad = 1.
\end{align*}
\end{IEEEproof}

%Construction \ref{con:debruijn} rely on certain assumptions on the parameters $n$ and $m$. Those assumptions were used in order to allow an easier description of the construction. However, a practical application doesn't have to be restricted to those assumptions in order to achieve an asymptotically optimal rate. For example, the number of blocks and the length of the block (both equal to $m$ in the construction), might differ by a constant number, and the rate of the code would still approach the optimal value.

\section{Conclusions}
\label{sec:conclusion}
We presented the framework for $(s,t,n)$-local rank modulation, and studied Gray codes for the most general case. The codes we present are asymptotically optimal.

Several questions remain open. For the case of $(1,2,n)$-LRM, a previous work describes asymptotically-optimal codes for which the weight of the codewords is constant and approaches $\frac{n}{2}$ \cite{EngLanSchBru10}. That property guarantees a bounded charge difference in any ``push-to-the-top'' operation. Constant-weight codes for the general case are still missing. Of more general interest is the study of codes that cover a constant fraction of the space.

%%%%%%%%%%%%%%%%%%%%%%%%%%%%%%%%%%%%%%%%%%%%%%%%%%%%%%%%%%%%%%%%%%%%%%%%
%%%%%%%%%%%%%%%%%%%%%%%%%%%%%%%%%%%%%%%%%%%%%%%%%%%%%%%%%%%%%%%%%%%%%%%%
%%%%%%%%%%%%%%%%%%%%%%%%%%%%%%%%%%%%%%%%%%%%%%%%%%%%%%%%%%%%%%%%%%%%%%%%
\bibliographystyle{IEEEtranS}
\bibliography{allbib}

% Generated by IEEEtranS.bst, version: 1.13 (2008/09/30)
\begin{thebibliography}{10}
\providecommand{\url}[1]{#1}
\csname url@samestyle\endcsname
\providecommand{\newblock}{\relax}
\providecommand{\bibinfo}[2]{#2}
\providecommand{\BIBentrySTDinterwordspacing}{\spaceskip=0pt\relax}
\providecommand{\BIBentryALTinterwordstretchfactor}{4}
\providecommand{\BIBentryALTinterwordspacing}{\spaceskip=\fontdimen2\font plus
\BIBentryALTinterwordstretchfactor\fontdimen3\font minus
  \fontdimen4\font\relax}
\providecommand{\BIBforeignlanguage}[2]{{%
\expandafter\ifx\csname l@#1\endcsname\relax
\typeout{** WARNING: IEEEtranS.bst: No hyphenation pattern has been}%
\typeout{** loaded for the language `#1'. Using the pattern for}%
\typeout{** the default language instead.}%
\else
\language=\csname l@#1\endcsname
\fi
#2}}
\providecommand{\BIBdecl}{\relax}
\BIBdecl

\bibitem{BerJelWol72}
T.~Berger, F.~Jelinek, and J.~K. Wolf, ``Permutation codes for sources,''
  \emph{IEEE Trans.~on Inform.~Theory}, vol. IT-18, no.~1, pp. 160--169, Jan.
  1972.

\bibitem{Bla74}
I.~F. Blake, ``Permutation codes for discrete channels,'' \emph{IEEE Trans.~on
  Inform.~Theory}, vol.~20, pp. 138--140, 1974.

\bibitem{BlaCohDez79}
I.~F. Blake, G.~Cohen, and M.~Deza, ``Coding with permutations,''
  \emph{Inform.~and Control}, vol.~43, pp. 1--19, 1979.

\bibitem{BohJiaBru07}
V.~Bohossian, A.~Jiang, and J.~Bruck, ``Buffer coding for asymmetric
  multi-level memory,'' in \emph{Proceedings of the 2007 IEEE International
  Symposium on Information Theory (ISIT2007), Nice, France}, Jun. 2007, pp.
  1186--1190.

\bibitem{ChaRee70}
H.~Chadwick and I.~Reed, ``The equivalence of rank permutation codes to a new
  class of binary codes,'' \emph{IEEE Trans.~on Inform.~Theory}, vol.~16,
  no.~5, pp. 640--641, 1970.

\bibitem{ChaKur69}
H.~D. Chadwick and L.~Kurz, ``Rank permutation group codes based on {K}endall's
  correlation statistic,'' \emph{IEEE Trans.~on Inform.~Theory}, vol. IT-15,
  no.~2, pp. 306--315, Mar. 1969.

\bibitem{CohDez77}
G.~Cohen and M.~Deza, ``Decoding of permutation codes,'' in \emph{Intl.~CNRS
  Colloquium, July, France}, 1977.

\bibitem{DezFra77}
M.~Deza and P.~Frankl, ``On maximal numbers of permutations with given maximal
  or minimal distance,'' \emph{J.~Combin.~Theory Ser.~A}, vol.~22, 1977.

\bibitem{EngLanSchBru10}
E.~{En Gad}, M.~Langberg, M.~Schwartz, and J.~Bruck, ``On a construction for
  constant-weight gray codes for local rank modulation,'' in \emph{Proceedings
  of the 2010 IEEE 26-th Convention of Electrical and Electronic Engineers in
  Israel (IEEEI2010), Eilat, Israel}, Nov. 2010, p. 996.

\bibitem{FerVinSwaBee05}
H.~C. Ferreira, A.~J.~H. Vinck, T.~G. Swart, and I.~de~Beer, ``Permutation
  trellis codes,'' \emph{IEEE Trans.~on Communications}, pp. 1782--1789, Nov.
  2005.

\bibitem{Gol67}
S.~W. Golomb, \emph{Shift Register Sequences}.\hskip 1em plus 0.5em minus
  0.4em\relax Holden-Day, San Francisco, 1967.

\bibitem{JiaBohBru07}
A.~Jiang, V.~Bohossian, and J.~Bruck, ``Floating codes for joint information
  storage in write asymmetric memories,'' in \emph{Proceedings of the 2007 IEEE
  International Symposium on Information Theory (ISIT2007), Nice, France}, Jun.
  2007, pp. 1166--1170.

\bibitem{JiaBru08}
A.~Jiang and J.~Bruck, ``Joint coding for flash memory storage,'' in
  \emph{Proceedings of the 2008 IEEE International Symposium on Information
  Theory (ISIT2008), Toronto, Canada}, Jul. 2008, pp. 1741--1745.

\bibitem{JiaLanSchBru09}
A.~Jiang, M.~Langberg, M.~Schwartz, and J.~Bruck, ``Universal rewriting in
  constrained memories,'' in \emph{Proceedings of the 2009 IEEE International
  Symposium on Information Theory (ISIT2009), Seoul, Korea}, Jun. 2009, pp.
  1219--1223.

\bibitem{JiaMatSchBru09}
A.~Jiang, R.~Mateescu, M.~Schwartz, and J.~Bruck, ``Rank modulation for flash
  memories,'' \emph{IEEE Trans.~on Inform.~Theory}, vol.~55, no.~6, pp.
  2659--2673, Jun. 2009.

\bibitem{JiaSchBru10}
A.~Jiang, M.~Schwartz, and J.~Bruck, ``Correcting charge-constrained errors in
  the rank-modulation scheme,'' \emph{IEEE Trans.~on Inform.~Theory}, vol.~56,
  no.~5, pp. 2112--2120, May 2010.

\bibitem{Lai88}
C.~A. Laisant, ``Sur la num\'{e}ration factorielle, application aux
  permutations,'' \emph{Bulletin de la Soci\'{e}t\'{e} Math\'{e}matique de
  France}, vol.~16, pp. 176--183, 1888.

\bibitem{Sch10}
M.~Schwartz, ``Constant-weight {G}ray codes for local rank modulation,'' in
  \emph{Proceedings of the 2010 IEEE International Symposium on Information
  Theory (ISIT2010), Austin, TX, U.S.A.}, Jun. 2010, pp. 869--873.

\bibitem{Sle65}
D.~Slepian, ``Permutation modulation,'' in \emph{Proc.~of the IEEE}, vol.~53,
  no.~3, 1965, pp. 228--236.

\bibitem{TamSch10}
I.~Tamo and M.~Schwartz, ``Correcting limited-magnitude errors in the
  rank-modulation scheme,'' \emph{IEEE Trans.~on Inform.~Theory}, vol.~56,
  no.~6, pp. 2551--2560, Jun. 2010.

\bibitem{VinHaeWad00}
H.~Vinck, J.~Haering, and T.~Wadayama, ``Coded {M-FSK} for power line
  communications,'' in \emph{Proceedings of the 2000 IEEE International
  Symposium on Information Theory (ISIT2000), Sorrento, Italy}, 2000, p. 137.

\bibitem{WanJiaBru09}
Z.~Wang, A.~Jiang, and J.~Bruck, ``On the capacity of bounded rank modulation
  for flash memories,'' in \emph{Proceedings of the 2009 IEEE International
  Symposium on Information Theory (ISIT2009), Seoul, Korea}, Jun. 2009, pp.
  1234--1238.

\bibitem{YaaSieWol08}
E.~Yaakobi, P.~H. Siegel, and J.~K. Wolf, ``Buffer codes for multi-level flash
  memory,'' in \emph{Proceedings of the 2008 IEEE International Symposium on
  Information Theory (ISIT2008), Toronto, Canada}, 2008, poster.

\end{thebibliography}

\end{document}